# Spectroscopic confirmation of a mature galaxy cluster at redshift two

J. P. Willis[1], R. E. A. Canning[2], E. S. Noordeh[2,3], S. W. Allen[2], A. L. King[2], A. Mantz[2,3], R. G. Morris[2,4], S. A. Stanford[5] & G. Brammer[6]

[1]Department of Physics and Astronomy, University of Victoria, Victoria, Canada.

[2]Kavli Institute for Particle Astrophysics and Cosmology, Stanford University, Stanford, CA, USA.

[3]Department of Physics, Stanford University, Stanford, CA, USA.

[4]SLAC National Accelerator Laboratory, Menlo Park, CA, USA.

[5]Department of Physics, University of California, Davis, CA, USA.

[6]Cosmic Dawn Centre, Niels Bohr Institute, University of Copenhagen, Copenhagen, Denmark.

*e-mail: jwillis@uvic.ca

**Galaxy clusters are the most massive virialized structures in the Universe and are formed through the gravitational accretion of matter over cosmic time[1]. The discovery[2] of an evolved galaxy cluster at redshift $z = 2$, corresponding to a look-back time of 10.4 billion years, provides an opportunity to study its properties. The galaxy cluster XLSSC 122 was originally detected as a faint, extended X-ray source in the XMM Large Scale Structure survey and was revealed to be coincident with a compact over-density of galaxies[2] with photometric redshifts of 1.9 ± 0.2. Subsequent observations[3] at millimetre wavelengths detected a Sunyaev–Zel'dovich decrement along the line of sight to XLSSC 122, thus confirming the existence of hot intracluster gas, while deep imaging spectroscopy from the European Space Agency's X-ray Multi-Mirror Mission (XMM-Newton) revealed[4] an extended, X-ray-bright gaseous atmosphere with a virial temperature of 60 million Kelvin, enriched with metals to the same extent as are local clusters. Here we report rest frame optical spectroscopic observations of XLSSC 122 and identify 37 member galaxies at a mean redshift of 1.98, corresponding to a look-back time of 10.4 billion years. We use photometry to determine a mean, dust-free stellar age of 2.98 billion years, indicating that star formation commenced in these galaxies at a mean redshift of 12, when the Universe was only 370 million years old. The full range of inferred formation redshifts, including the effects of dust, covers the interval from 7 to 13. These observations confirm that XLSSC 122 is a remarkably mature galaxy cluster with both evolved stellar populations in the member galaxies and a hot, metal-rich gas composing the intracluster medium**.

To further our understanding of this galaxy cluster, particularly the properties of its member galaxies, we undertook a series of observations of XLSSC 122 with the Hubble Space Telescope (HST) Wide Field Camera 3 (WFC3). We obtained images of the cluster in two wavebands, F105W and F140W, and performed low-spectral-resolution slitless spectroscopy using the G141 grism (see Methods). These observations cover the observed frame wavelength interval 1.0–1.7 μm, corresponding to an interval of 0.33 μm to 0.57 μm in the rest-frame of a galaxy at redshift $z = 2$. Figure 1 displays the F140W image of XLSSC 122 and shows a compact cluster of galaxies associated with the extended X-ray-emitting region.

We extracted one-dimensional spectra of all galaxies identified within the dispersed G141 grism image of the field (see Methods) and computed redshifts using a galaxy template-fitting algorithm with redshift as a free parameter. Figure 2 displays the histogram of galaxy redshifts in the field of XLSSC 122 over the restricted interval $1.9 < z < 2.05$. Inspection of this interval reveals a primary peak at $z = 1.98$ associated with the central, red galaxies closest to the X-ray peak, and a secondary redshift peak at $z = 1.93$ associated with a mixture of red and blue galaxies, located at larger projected cluster-centric distances (see Fig. 1). The line-of-sight separation between $z = 1.93$ and $z = 1.98$ is 76 co-moving megaparsecs, far larger than the size of the XLSSC 122 cluster, and the two structures are therefore physically distinct. As outlined in the Methods, we identify 37 galaxies as being members of the cluster and a further 13 galaxies identified as members of the foreground structure.

We performed photometry of all galaxies within the HST field of view in both the F105W and F140W images (see Methods) and summarize this information in Fig. 3. The galaxies identified at $1.9 < z < 2.05$ form a clear bimodal distribution in colour with a well populated sequence of red galaxies (corresponding to larger values of F105W-F140W) clearly separated from a broader distribution of blue galaxies.

Interpreting this red sequence as representing a restricted locus of star-formation histories, Bower, Lucey and Ellis[5] employed B-V (the astronomical magnitude difference between a blue and a visual filter) photometry of red-sequence galaxies in the Virgo and Coma clusters to constrain the dispersion of stellar ages in their member galaxies. The F105W and F140W photometry obtained for XLSSC 122 at $z = 2$ spans almost exactly the age-sensitive break feature at wavelength 4,000 Å in the member galaxy rest-frame spectral energy distributions (SEDs) and permits a similar analysis.

We therefore employed the F105W and F140W photometry of red-sequence galaxies to constrain the posterior distributions of luminosity-weighted stellar age and stellar mass for a set of synthetic stellar population models (see Methods). Computing the product of these posterior distributions generates a mean posterior on the luminosity-weighted stellar age of the red-sequence cluster members (Fig. 4), the details of which are presented in Table 1.

**Table 1 The mean luminosity-weighted stellar ages of red-sequence cluster galaxies**

| SED model $A_v$ | Mean $t_w$ (Gyr) | Mean formation redshift (spread) |
|---|---|---|
| 0.0 | 2.98 ± 0.05 | 12.0 (10.9-13.3) |
| 0.3 | 2.77 ± 0.13 | 8.7 (8.3-10.4) |
| 0.5 | 2.63 ± 0.11 | 7.4 (6.6-8.3) |

The table lists the mean and standard deviation of the computed luminosity-weighted stellar age $t_w$ for SED models of specified $A_V$. Corresponding values of the mean formation redshift and spread (from standard deviation) are computed for the assumed cosmological model.

Our analysis assumes that the tight correlation of colour on the red sequence arises from scatter in age at fixed metallicity and internal dust absorption. Assuming no dust absorption ($A_V = 0.0$) we determine a mean red-sequence luminosity-weighted stellar age of 2.98 billion years (Gyr), corresponding to a redshift marking the onset of star formation of 12. This value is consistent with the inferred formation redshifts of the earliest observations of star formation in the Universe[6]. We also consider dust absorption characterized by $A_V = 0.3$ and $A_V = 0.5$, which generate lower mean stellar ages and greater dispersion of the mean age. Despite the uncertainties that govern a number of the assumptions in such stellar population analyses, the main conclusion of this analysis is that red-sequence galaxies in XLSSC 122 are composed of stars of uniformly old age. When combined with the already large look-back time to this cluster, it is clear that star formation occurred in these galaxies in a coordinated manner at early times.

Analysis of the X-ray-emitting gas in XLSSC 122 provides an alternative perspective on the formation history of the galaxy cluster as a whole. Using standard theory[7], one may combine the X-ray gas temperature ($k_B T = 5$ keV, where $k_B$ is the Boltzmann constant) and an estimate of the virial radius of the cluster ($r_{200} = 440$ kpc, ref. [4]) to obtain a sound-crossing time for XLCCS 122 of $3.3 \times 10^8$ yr. Hydrodynamical simulations of the gas physics in a forming cluster indicate that structures typically achieve virial equilibrium following a minimum of 2 to 3 sound-crossing timescales[8]. This indicates that XLSSC 122 is unlikely to have assembled earlier than 1 Gyr before the epoch of observation, equivalent to a redshift of 2.8. Although this argument does not place an upper limit on the elapsed time between assembly and virialization, the assembly of a cluster of mass equal to XLSSC 122 at redshifts greater than 3 appears unlikely[9,10]. It is clear, therefore, even allowing for the uncertainty present in these estimates, that coordinated star formation in the member galaxies located in XLSSC 122 preceded the assembly of the cluster environment.

Computer simulations of the accretion history of massive, gravitationally bound halos in an expanding Universe indicate that it is likely that XLSSC 122 will evolve with time into a present-day galaxy cluster comparable in mass to that of Coma, that is, about $1 \times 10^{15}$ solar masses[9–11]. Although caution is required in both the interpretation of the scatter in the accretion histories of halos of fixed total mass and the more subtle point of whether XLSSC 122 represents a typical galaxy cluster at $z = 2$ or is perhaps an extreme case, the conclusion remains robust that this system will continue to grow in mass until it becomes a massive galaxy cluster in the present-day Universe.

The recent discovery of SPT2349-56, a massive proto-cluster of galaxies at a redshift of 4.3 (ref. [12]), provides a further, tantalising glimpse of the kind of structure from which XLSSC 122 may have evolved. The same structure growth simulations that predict the future evolution of massive halos can also be used to infer their likely past accretion histories. Even taking into account the caveats expressed above, such simulations indicate that structures such as SPT2349, XLSSC 122 and Coma may represent similar clusters viewed at very different cosmic epochs. From such studies we are beginning to achieve a coherent view of the formation and evolution of the largest gravitationally bound structures in the Universe.

**Acknowledgements** We acknowledge the builders of the XMM-LSS and XXL surveys, on whose work this paper is based. This work is based on observations made with the NASA/ESA Hubble Space Telescope, obtained at the Space Telescope Science Institute, which is operated by the Association of Universities for Research in Astronomy, Inc., under NASA contract NAS5-26555. These observations are associated with programme number 15267. J.P.W. and E.S.N. acknowledge support from NSERC. R.E.A.C., E.S.N., S.W.A, A.L.K., A.M. and R.G.M. acknowledge support from NASA grant number HST-GO-15267.002-A. The Cosmic Dawn Center is funded by the Danish National Research Foundation.

**Author contributions** J.P.W., R.E.A.C. and E.S.N. analysed the data and wrote the paper. S.W.A., A.L.K., A.M., R.G.M., S.A.S. and G.B. provided guidance on the analysis and commented on the paper.

## Methods

We assume a Lambda cold dark matter (CDM) cosmological model described by the parameters $\Omega_{\mathrm{M}} = 0.286$, $\Omega_{\Lambda} = 0.714$, $H_0 = 69.6$ km s$^{-1}$ Mpc$^{-1}$ (ref. [13]). The present-day age of the Universe in this model is 13.72 Gyr. All magnitude information is presented using the AB system.

The HST observations were obtained between 4 November 2017 and 13 January 2018 and comprised one orbit in F105W and 12 orbits in the F140W+G141 filter and grism combination. The 12 F140W+G141 orbits were split into three orbits at each of four orientations using an ABB BBA pattern for exposures in each orbit. The total exposure times in F105W, F140W and G141 were, respectively, 2,612 s, 5,171 s and 26,541 s.

The imaging and spectroscopic observations were reduced with Grizli version 0.3.0 (ref. [14]). Raw HST data products were processed by applying standard image-calibration techniques with additional corrections applied for variable backgrounds (the HST reduction pipeline calwf3 assumes a constant background not appropriate for WFC3 infrared observations) and to mask artefacts such as satellite trail features[15,16]. Relative and absolute astrometric registration were achieved by aligning to reference sources in the Sloan Digital Sky Survey. The final steps included flat fielding and master background subtraction for both the direct and grism images, and drizzling of the individual data frames to produce stacked images.

Reduced data were processed with SExtractor (version 2.5.0) to generate photometric catalogues. The F140W image was processed using standard WFC3 zero point information with the gain parameter set to the image exposure time. Source detection used a pixel-based inverse variance weighting (pipeline IVM file in SExtractor), whereas source photometry employed a root mean square (pipeline RMS file in SExtractor) variation per pixel weight. The F105W image was processed employing the SExtractor two-image mode with the F140W image used as the detection image. Source fluxes and AB magnitudes were computed within two apertures: a 0.8-arcsecond circular aperture[17,18] and an elliptical aperture based upon the Kron radius (a statistical moment computed from the surface brightness distribution in each object) with the Kron factor set to $k = 0.8$ to avoid excessive source blending in the central cluster regions. Sources with a half-light radius of <0.22 arcseconds were classified as stellar. In the following analysis we consider sources brighter than F140W = 25.5, corresponding to an image signal-to-noise ratio (SNR) >10.

Spectral extraction from the G141 images employed the F140W segmentation map produced by SExtractor (see above) to identify undispersed source positions. These source positions were then employed to construct a full field contamination model of each G141 image. The contamination model initially assumes a spectrally flat continuum for all sources brighter than 25th magnitude in F140W. This provides a first-pass estimate of those pixels contaminated by spectra from more than one source. Spectral traces, represented by 2nd-order polynomial functions, were fitted to all of the above bright sources in each exposure at each orientation. Extracted spectra for these sources were then employed to compute a second-pass contamination model. The model was further refined for 26 bright objects, which were identified as contaminating the spectra of bright red-sequence galaxies. In these cases synthetic stellar population models were fitted to the contaminating spectra and these updated spectral models were propagated to the global contamination model. Employing the above procedures, and with the G141 observations split into four orientations, we were able to obtain a satisfactory contamination model for most sources even in such a densely packed field.

Two-dimensional spectra were extracted separately for each G141 exposure and resulted in a maximum of 48 spectral extractions per source. These spectra were optimally extracted[19] and simultaneously fitted with a suite of galaxy templates[20]. The templates were stepped in redshift over a coarse ($\Delta z = 0.01$) grid from $z = 0.2$–4.0 and subsequently refitted over a fine grid ($\Delta z = 0.0004$) in redshift around peaks in the probability distribution function.

We define the SNR of each spectrum as the average spectral flux per pixel divided by the pipeline-computed noise per pixel integrated over the wavelength interval 1.3–1.55 μm. A galaxy of brightness $F140W_{Kron} = 24$ typically generates a spectrum of SNR = 5 with a scatter consistent with random noise. We inspected visually all spectra displaying a spectral SNR >2 to assess the reliability of the fitted redshift and template model. We concluded that all spectra displaying SNR ≥ 5 possess a visually reliable redshift measurement and consequently we employ $F140W_{Kron} = 24$ as the galaxy brightness corresponding to our spectroscopic completeness limit. Furthermore, we determined that sources with visually identified emission lines possess a reliable redshift to a limit of SNR ≥ 3, corresponding to a brightness $F140W_{Kron} = 24.5$, which we adopt for our spectroscopic completeness limit for emission line sources. Extended Data Fig. 1 shows two examples of extracted grism spectra.

Galaxy membership of the $z = 1.98$ cluster was defined according to a number of criteria that we describe below. We define 'gold' members as those displaying $F140W_{Kron} = 24$ (24.5 for emission line sources) and $P_{mem} > 0.5$ where $P_{mem}$ is defined as the integral of the redshift probability distribution function for each galaxy over the interval $1.96 < z < 2.00$. This interval corresponds to $z_{cluster} \pm 3\sigma_z$ where $\sigma_z$ is the observed frame velocity dispersion of a 5-keV galaxy cluster expressed in redshift space[21]. There are 33 galaxies in this class (of which four are emission line sources with $24 < F140W_{Kron} \leq 24.5$). We compute the redshift of XLSSC 122 as the unweighted mean of the 'gold' cluster member redshifts. The redshift is $z = 1.978 \pm 0.010$. We define 'silver' members as those displaying $0.1 < P_{mem} < 0.5$—a change that adds four new members (one of which is on the red sequence)—for a total of 37. Finally, we create an additional class to identify members of the z = 1.93 foreground structure as those displaying $P'_{mem} > 0.5$, where $P'_{mem}$ is defined as the integral of the redshift probability distribution function for each galaxy over the interval $1.91 < z < 1.95$ with the same brightness limits as before. There are 13 galaxies in this class. We compute the redshift of this structure as the unweighted mean redshift of these 13 galaxies. The resulting structure redshift is $z = 1.934 \pm 0.007$. This analysis therefore identifies a total of 50 galaxies that are members of either XLSSC 122 or the $z = 1.93$ structure (see Extended Data Table 1 and Fig. 1 for these members plotted on the greyscale HST/WFC3 image).

Figure 3 shows the colour–magnitude diagram for all galaxies identified within the HST field. We identify a total of 30 red-sequence members according to $1.15 < F105W_{ap}$-$F140W_{ap} < 1.65$ and $F140W_{Kron} < 24$. Of these, 19 are defined as 'gold' cluster members as described above. Of the remaining 11 galaxies, one is a silver cluster member, one is located at $z \approx 2$ with a relatively broad redshift probability distribution function, five are located within the $z = 1.93$ structure and four are located at $z > 2$ yet display spectra affected by source confusion and contamination. We restrict our subsequent red-sequence analysis to the 19 'gold' cluster members. An unweighted, linear, least-squares fit to the red-sequence members generates the angled solid line shown in Fig. 3. The root-mean-square deviation in colour about this line normalized by the photometric error is 1.72, that is, the observed scatter is 72% larger than expected from the computed colour errors.

At redshifts $z < 1$, the dominant populations of evolved, red galaxies are interpreted to be the result of the prompt suppression of star formation within galaxies accreting into the cluster environment[22]. The details of this process, euphemistically referred to as 'quenching', remain uncertain, with likely physical scenarios including the ram pressure stripping of gas from galaxies falling through the hot, X-ray-emitting, intra-cluster medium[23,24]. The exact mass scale at which quenching occurs is also debatable, with uncertainty as to whether the suppression of star formation occurs as galaxies are accreted into less massive groups before encountering more massive clusters[25]. With a quenched fraction of 0.51 ± 0.14 at a look-back time of 10.4 Gyr for XLSSC 122, it is clear that the physical processes involved in quenching were established at an even earlier cosmic epoch.

We employ the F105W and F140W photometry of red-sequence galaxies to constrain the stellar age, star formation rate and stellar mass of a set of synthetic stellar population models. The analysis presented in this paper intentionally follows that performed by Andreon *et al.*[26] and Newman *et al.*[27] of galaxies within the cluster JKCS 041 at $z = 1.8$. This was done in order to allow as direct a comparison as possible between galaxy populations in two high redshift clusters, albeit observed in different photometric filters.

We employ a grid of simple stellar population models[28] to generate synthetic F105W and F140W photometry. The likelihood of the model photometry given the data is expressed as $L = \exp(-\chi^2/2)$, where:

$$\chi^2 = \sum_i \left( \frac{D_i - M_i}{\sigma_i} \right)^2$$

and $D_i$ represents the measured apparent magnitude in the $i$th filter, $M_i$ is the apparent magnitude computed from the stellar population model and $\sigma_i$ is the uncertainty in the measured magnitude. The stellar population models are characterized by an exponentially declining burst of star formation where the star formation rate SFR $\propto \exp(-t/\tau)$. The variable $t$ denotes the time since the burst commenced and $\tau$ is the e-folding time. Models are further characterized by a Salpeter[29] initial mass function and solar metallicity. Gas lost during stellar evolution is not recycled and the effects of dust are included by applying a Calzetti attenuation law[30] parameterized by the extinction parameter $A_V$ at 5,500 Å. The stellar population model photometry is normalized per unit stellar mass and is scaled by a total stellar mass variable, $M_{star}$.

The stellar population model grid spans $8 < \log[t \text{ (yr)}] < 9.7$ and $8 < \log[\tau \text{ (yr)}] < 9.7$. We compute posterior distributions in $\log t$, $\log\tau$ and $\log M_{\rm star}$, employing a Markov chain Monte Carlo algorithm and assuming flat priors. We do not explore the $A_{\rm V}$ posterior explicitly at this stage. Instead we compute the posterior distributions of the above variables at three explicit values of $A_{\rm V}$ (0.0, 0.3 and 0.5). Finally, we compute the average luminosity-weighted stellar population age, following refs. [26,31], as:

$$t_{\rm w} = \frac{t}{1 - e^{-t/\tau}} - \tau$$

The posterior distributions of $t_{\rm w}$ and $M_{\rm star}$ are displayed in Extended Data Fig. 2 for the 19 'gold' cluster members. Extended Data Fig. 3 displays the one-dimensional posterior distribution in $t_{\rm w}$ for each cluster member having marginalized over $M_{\rm star}$. In addition, Fig. 4 compares the average $t_{\rm w}$ posterior for all cluster members, computed as the product of individual posteriors, for each of the three dust models, $A_{\rm V} = 0.0$, 0.3 and 0.5. Values of mean luminosity-weighted stellar age and standard deviation are listed in Table 1. For the canonical model employing zero dust absorption the mean stellar age of 2.98 Gyr at $z = 1.98$ corresponds to a mean star formation redshift of $z = 12.0$.

The spread of stellar age values in XLSSC 122 overlaps with those determined for the galaxy cluster JKCS 041 at $z = 1.803$ (ref. [26]), yet the mean stellar age in XLSSC 122 is older, even though the Universe is 0.32 Gyr younger at $z = 1.98$ compared to $z = 1.8$. Although this comparison employs the same analysis methodology, the two clusters are observed using different photometric filters, while the clusters themselves may represent very different structures. It is instructive therefore to further compare our results to the study of Strazzullo *et al.*[32] who also analysed HST WFC3 F105W and F140W photometry for the $z = 2$ cluster CL1449+0856[33]. Applying a stellar population model characterized by a short (0.25 Gyr) burst of metal-rich (150% the solar value) star formation, they obtain a typical formation redshift of 3 to 5 for galaxies of similar colour and redshift to those analysed in XLSSC 122. Applying a similar model to the data for XLSSC 122, we obtain a typical stellar population age of 1.4 Gyr, corresponding to formation redshift of 3.3, in agreement with ref. [32]. Ultimately, we consider the assumption of a short, metal-rich burst of star formation to be unnecessarily restrictive given the considerable uncertainty regarding the exact physical state of these high-redshift stellar populations and adopt a more flexible approach as outlined in this paper. Overall however, the comparison is instructive because it highlights the key influence of the assumptions governing the stellar population model upon the inferred formation redshift of the luminosity weighted stellar content of the cluster member galaxies. The acquisition of further data, in particular concerning the dust and metal content of the member galaxies of these high-redshift clusters provides a clear observational route to resolving such issues. We therefore emphasize in conclusion that the results of such stellar population modelling, when based upon broad-band photometry, are most conservatively interpreted as indicating the range of physically reasonable input parameters and not as indicating a definitive physical state of the stellar population.

**Data availability**

All HST data presented in this paper are publicly available at the Hubble Legacy Archive (https://hla.stsci.edu/). The programme number is 15267.

## References

1. Kravtsov, A. V. & Borgani, S. Formation of galaxy clusters. *Annu. Rev. Astron. Astrophys.* **50,** 353–409 (2012).

2. Willis, J. P. *et al.* Distant galaxy clusters in the XMM Large Scale Structure survey. *Mon. Not. R. Astron. Soc.* **430,** 134–156 (2013).

3. Mantz, A. B. *et al.* The XXL survey. V. Detection of the Sunyaev-Zel'dovich effect of the redshift 1.9 galaxy cluster XLSSU J021744.1–034536 with CARMA. *Astrophys. J.* **794,** 157 (2014).

4. Mantz, A. B. *et al.* The XXL survey. XVII. X-ray and Sunyaev-Zel'dovich properties of the redshift 2.0 galaxy cluster XLSSC 122. *Astron. Astrophys.* **620,** A2 (2018).

5. Bower, R. G., Lucey, J. R. & Ellis, R. S. Precision photometry of early-type galaxies in the Coma and Virgo clusters: a test of the universality of the colour-magnitude relation. II. Analysis. *Mon. Not. R. Astron. Soc.* **254,** 601–613 (1992).

6. Hashimoto, T. *et al.* The onset of star formation 250 million years after the Big Bang. *Nature* **557,** 392–395 (2018).

7. Sarazin, C. L. *X-Ray Emission From Galaxy Clusters* (Cambridge Astrophysics Series, Cambridge University Press, 1988).

8. Roettiger, K., Stone, J. M. & Mushotzky, R. F. Anatomy of a merger: a numerical model of A754. *Astrophys. J.* **493,** 62–72 (1998).

9. Chiang, Y.-K., Overzier, R. A. & Gebhardt, K. Ancient light from young cosmic cities: physical and observational signatures of galaxy proto-clusters. *Astrophys. J.* **779,** 127 (2013).

10. Harrison, I. & Hotchkiss, S. A consistent approach to falsifying ΛCDM with rare galaxy clusters. *J. Cosmol. Astropart. Phys.* **7**, 022 (2013).

11. Gavazzi, R. *et al.* A weak lensing study of the Coma cluster. *Astron. Astrophys.* **498,** L33–L36 (2009).

12. Miller, T. B. *et al.* A massive core for a cluster of galaxies at a redshift of 4.3. *Nature* **556,** 469–472 (2018).

13. Bennett, C. L., Larson, D., Weiland, J. L. & Hinshaw, G. The 1% concordance Hubble constant. *Astrophys. J.* **794,** 135 (2014).

14. Brammer, G. GRIZLI: Grism redshift and line analysis software. *Astrophys. Source Code Library* record ascl:1905.001 (2019).

15. Brammer, G., Pirzkal, N., McCullough, P. & MacKenty, J. *Time-Varying Excess Earth-Glow Backgrounds In The WFC3/IR Channel.* Instrument Science Report WFC3 2014-03 **(**Space Telescope Science Institute 2014).

16. Brammer, G. *Reprocessing WFC3/IR Exposures Affected by Time-Variable Backgrounds.* Instrument Science Report WFC3 2016-16 (Space Telescope Science Institute 2016).

17. Stanford, S. A. *et al.* IDCS J1426.5+3508: discovery of a massive, infrared-selected galaxy cluster at z = 1.75. *Astrophys. J.* **753,** 164 (2012).

18. Kron, R. G. Photometry of a complete sample of faint galaxies. *Astrophys. J. Suppl. Ser.* **43,** 305–325 (1980).

19. Horne, K. An optimal extraction algorithm for CCD spectroscopy. *Publ. Astron. Soc. Pacif.* **98,** 609–617 (1986).

20. Brammer, G. B., van Dokkum, P. G. & Coppi, P. EAZY: A fast, public photometric redshift code. *Astrophys. J.* **686,** 1503–1513 (2008).

21. Ruel, J. *et al.* Optical spectroscopy and velocity dispersions of galaxy clusters from the SPT-SZ Survey. *Astrophys. J.* **792,** 45 (2014).

22. Stanford, S. A., Eisenhardt, P. R. & Dickinson, M. The evolution of early-type galaxies in distant clusters. *Astrophys. J.* **492,** 461–479 (1998).

23. Gunn, J. E. & Gott, J. R. III. On the infall of matter into clusters of galaxies and some effects on their evolution. *Astrophys. J.* **176,** 1–19 (1972).

24. Butcher, H. & Oemler, A. Jr The evolution of galaxies in clusters. II. The galaxy content of nearby clusters. *Astrophys. J.* **226,** 559–565 (1978).

25. Jung, S. *et al.* On the origin of gas-poor galaxies in galaxy clusters using cosmological hydrodynamic simulations. *Astrophys. J.* **865,** 156 (2018).

26. Andreon, S. *et al.* JKCS 041: a Coma cluster progenitor at z = 1.803. *Astron. Astrophys.* **565,** A120 (2014).

27. Newman, A. B. *et al.* Spectroscopic confirmation of the rich z = 1.80 galaxy cluster JKCS 041 using the WFC3 grism: environmental trends in the ages and structure of quiescent galaxies. *Astrophys. J.* **788,** 51 (2014).

28. Bruzual, G. & Charlot, S. Stellar population synthesis at the resolution of 2003. *Mon. Not. R. Astron. Soc.* **344,** 1000–1028 (2003).

29. Salpeter, E. E. The luminosity function and stellar evolution. *Astrophys. J.* **121,** 161–167 (1955).

30. Calzetti, D. *et al.* The dust content and opacity of actively star-forming galaxies. *Astrophys. J.* **533,** 682–695 (2000).

31. Longhetti, M. *et al.* Dating the stellar population in massive early-type galaxies at z~1.5. *Mon. Not. R. Astron. Soc.* **361,** 897–906 (2005).

32. Strazzullo, V. *et al.* The red sequence at birth in the galaxy cluster Cl J1449+0856 at z = 2. *Astrophys. J.* **833,** L20 (2016).

33. Gobat, R. *et al.* WFC3 GRISM confirmation of the distant cluster Cl J1449+0856 at <z> = 2.00: quiescent and star-forming galaxy populations. *Astrophys. J.* **776,** 9 (2013).

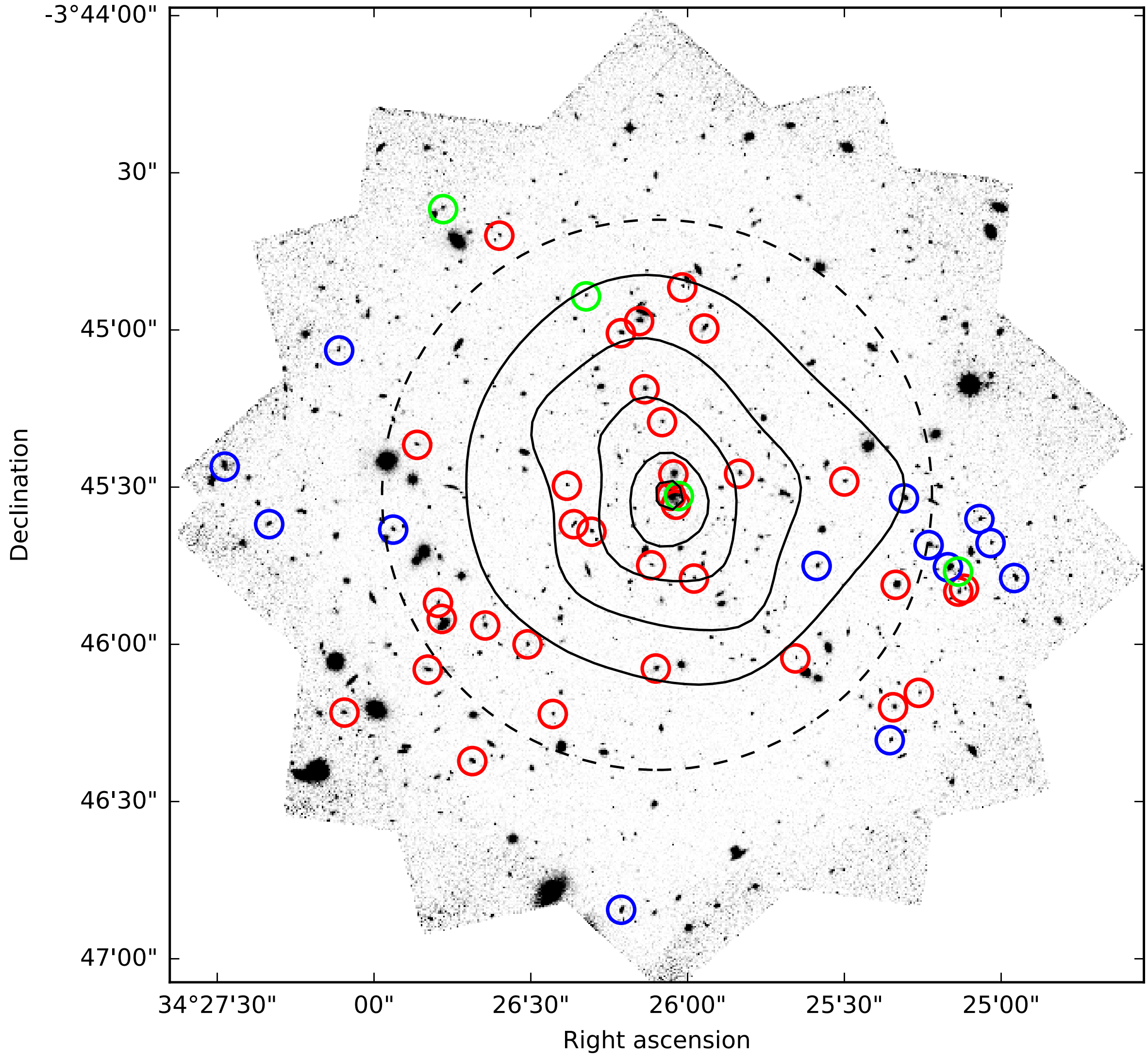

**Fig. 1 HST image of the galaxy cluster XLSSC 122.** The greyscale is the F140W image. Contours display X-ray emission corresponding to the 100-ks XMM-Newton image presented in ref. [4]. The dashed circle is drawn with a radius equal to the measured value of $r_{500}$ (the radius within which the average matter density is 500 times the critical density of the Universe). Spectroscopic 'gold' and 'silver' members (see Methods) of the $z = 1.98$ cluster are indicated by red and green circles, respectively. Members of the $z = 1.93$ foreground structure are indicated by blue circles. See text for further details.

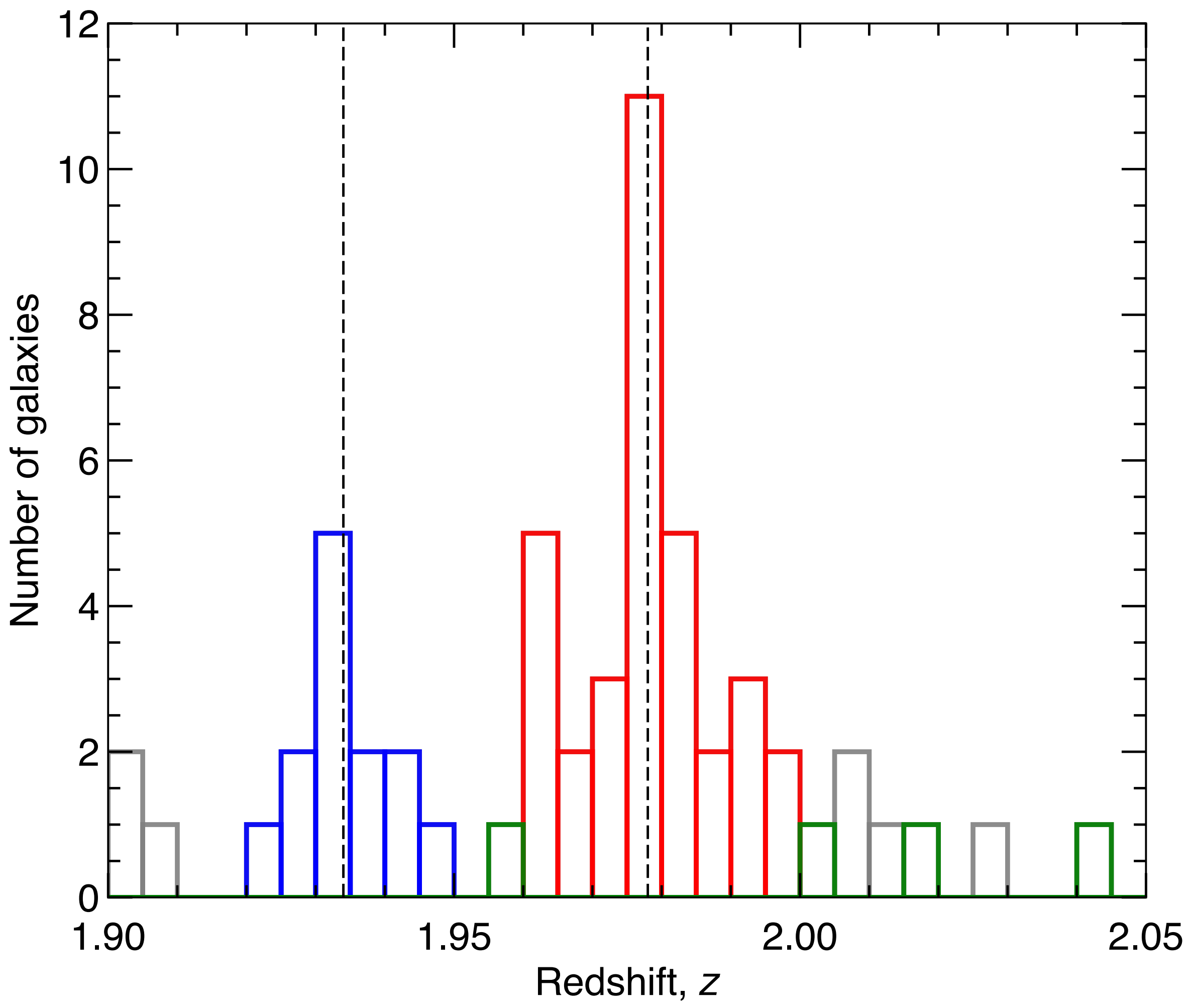

**Fig. 2 Redshift histogram of all galaxies along the line of sight to XLSSC 122.** The histogram considers galaxies satisfying the magnitude measurement $F140W_{Kron} < 24$. Galaxies classified as 'gold' members of the $z = 1.98$ cluster are shown in red, 'silver' members are shown in green and members of the $z = 1.93$ structure are shown in blue. Galaxies not classified as a member of either the $z = 1.98$ cluster or the $z = 1.93$ structure are shown in grey. The vertical dashed lines show the unweighted mean redshift of both the cluster and the foreground structure (see text for further details).

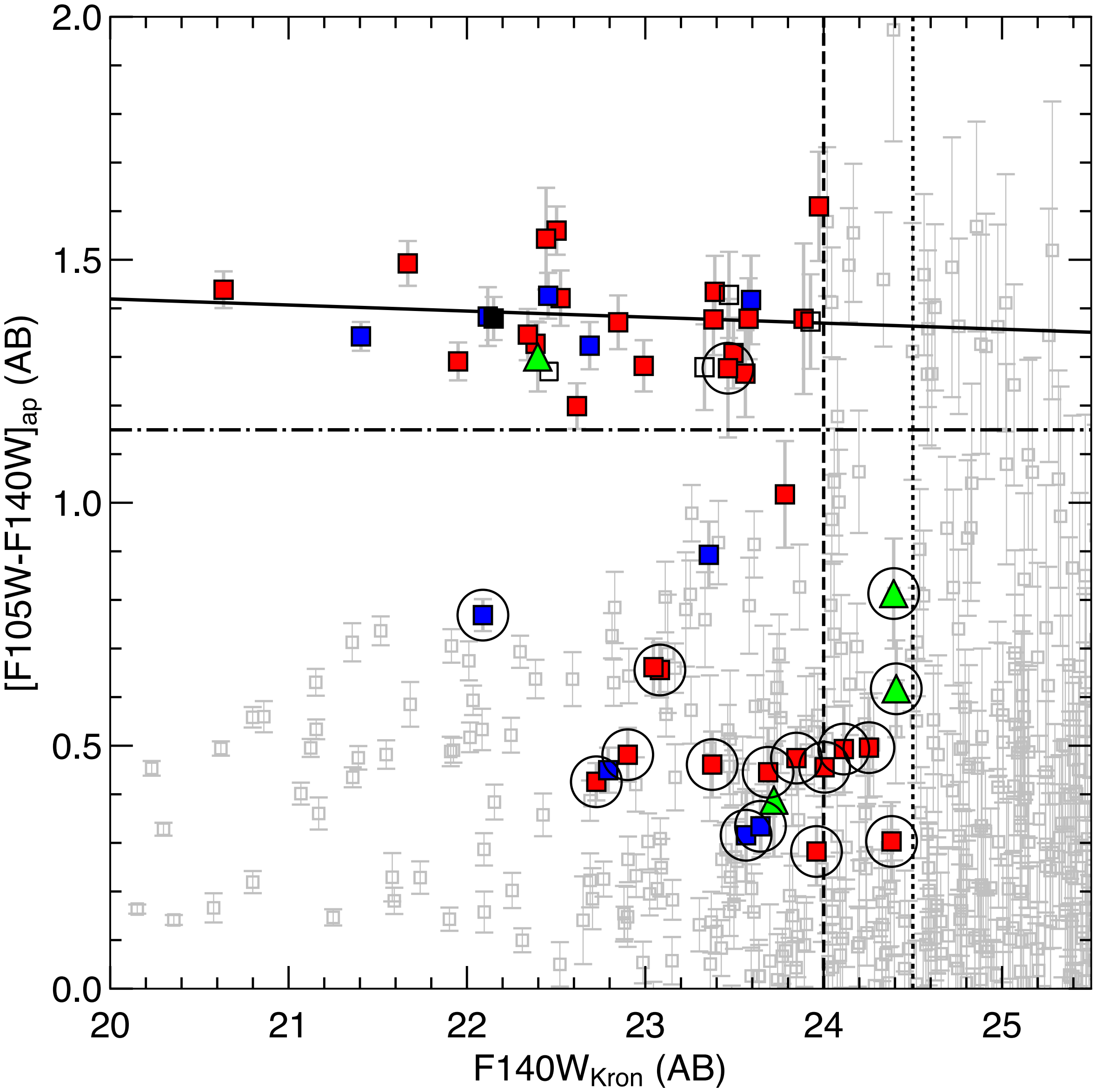

**Fig. 3 Colour–magnitude diagram of all galaxies within the HST/WFC3 field of view.** Spectroscopically confirmed $z = 1.98$ 'gold' and 'silver' cluster members are indicated as red squares and green triangles, respectively. Members of the $z = 1.93$ structure are indicated as blue squares. Galaxies at $z \approx 2$ yet which are not formal cluster members are shown as solid black squares, whereas potentially contaminated or confused spectroscopic sources are shown as open black squares. Galaxies with visually classified emission lines are marked using black circles (only $z = 1.98$ and $z = 1.93$ are marked in this manner). All other galaxies in the field are indicated by grey squares. Error bars indicate the 1-sigma measurement uncertainty. The spectroscopic completeness limits of $F140W_{Kron} = 24$ and 24.5 are indicated by the vertical dashed and dotted lines. The horizontal dot-dashed line shows the lower colour limit for a source to be considered on the cluster's red sequence. The angled solid line indicates a simple least-squares fit to the colour–magnitude relation for red-sequence cluster members.

Subscript "ap" indicates that the magnitudes of these objects are measured within an aperture of fixed angular size (as mentioned in methods) as opposed to a flexible aperture, referred to as "Kron".

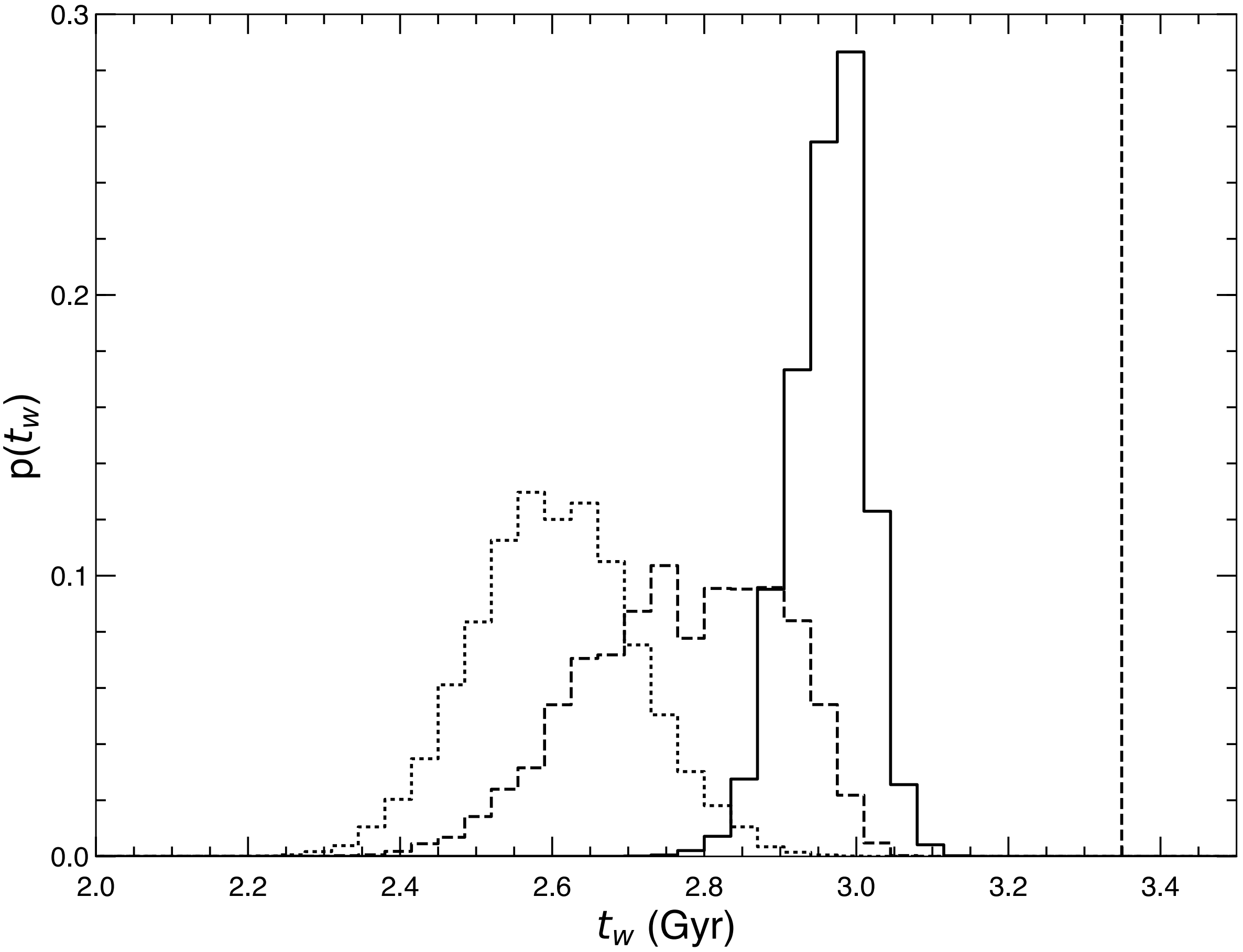

**Fig. 4 The luminosity-weighted age distribution of stars within red-sequence cluster galaxies.** The lines depict mean $t_w$ posteriors for 19 'gold' $z = 1.98$ cluster red-sequence members for each of the three SED models characterized by $A_V = 0.0$ (solid), $A_V = 0.3$ (dashed) and $A_V = 0.5$ (dotted). The vertical dashed line indicates the age of the Universe at $z = 1.98$ for the assumed cosmological model.

| ID | Right Ascension (deg.) | Decl. (deg.) | Magnitude | Colour | Redshift | Notes |
|---|---|---|---|---|---|---|
| 526 | 34.43422 | -3.75880 | 20.64 | 1.44 | 1.980 | G |
| 451 | 34.42228 | -3.76351 | 21.95 | 1.29 | 1.981 | G |
| 657 | 34.43410 | -3.75766 | 21.67 | 1.49 | 1.983 | G |
| 1032 | 34.43245 | -3.74992 | 22.38 | 1.33 | 1.982 | G |
| 295 | 34.43503 | -3.76795 | 22.50 | 1.56 | 1.987 | G |
| 917 | 34.43563 | -3.75314 | 22.73 | 0.43 | 1.963 | G |
| 298 | 34.44715 | -3.76801 | 22.52 | 1.42 | 1.993 | G |
| 1050 | 34.43689 | -3.75017 | 22.85 | 1.37 | 1.977 | G |
| 1064 | 34.43592 | -3.74954 | 22.34 | 1.35 | 1.988 | G |
| 606 | 34.43845 | -3.76070 | 22.99 | 1.28 | 1.966 | G |
| 240 | 34.42242 | -3.77000 | 22.62 | 1.20 | 1.977 | G |
| 845 | 34.43470 | -3.75489 | 23.38 | 1.38 | 1.979 | G |
| 372 | 34.44410 | -3.76567 | 23.08 | 0.66 | 1.963 | G |
| 734 | 34.42501 | -3.75803 | 23.39 | 1.43 | 1.996 | G |
| 1220 | 34.44335 | -3.74500 | 23.49 | 1.31 | 1.976 | G |
| 345 | 34.44185 | -3.76667 | 23.56 | 1.27 | 1.991 | G |
| 145 | 34.44478 | -3.77286 | 22.90 | 0.48 | 1.981 | G |
| 493 | 34.43300 | -3.76318 | 23.58 | 1.38 | 1.962 | G |
| 603 | 34.43939 | -3.76030 | 23.38 | 0.46 | 1.979 | G |
| 1141 | 34.43362 | -3.74775 | 23.85 | 0.47 | 1.963 | G |
| 402 | 34.44641 | -3.76532 | 23.05 | 0.66 | 1.972 | G |
| 730 | 34.43975 | -3.75826 | 23.97 | 1.61 | 1.993 | G |
| 649 | 34.43396 | -3.75927 | 22.44 | 1.54 | 1.997 | G |
| 726 | 34.43060 | -3.75762 | 23.46 | 1.28 | 1.969 | G |
| 452 | 34.41895 | -3.76387 | 23.69 | 0.45 | 1.971 | G |
| 806 | 34.44771 | -3.75609 | 23.78 | 1.02 | 1.981 | G |
| 236 | 34.45158 | -3.77029 | 23.02 | N/A | 1.977 | G |
| 547 | 34.43527 | -3.76248 | 23.96 | 0.28 | 1.963 | G |
| 428 | 34.44661 | -3.76447 | 23.89 | 1.38 | 1.977 | G |
| 466 | 34.41865 | -3.76372 | 24.11 | 0.49 | 1.979 | GE |
| 229 | 34.44051 | -3.77036 | 24.01 | 0.46 | 1.978 | GE |
| 329 | 34.42761 | -3.76741 | 24.38 | 0.30 | 1.972 | GE |
| 263 | 34.42106 | -3.76924 | 24.25 | 0.50 | 1.976 | GE |
| 642 | 34.43380 | -3.75881 | 22.40 | 1.30 | 2.041 | S |
| 1253 | 34.44633 | -3.74360 | 23.72 | 0.39 | 2.018 | S |
| 1125 | 34.43874 | -3.74822 | 24.41 | 0.62 | 2.000 | SE |
| 522 | 34.41896 | -3.76281 | 24.39 | 0.81 | 1.959 | SE |
| 462 | 34.41950 | -3.76258 | 21.41 | 1.34 | 1.930 | F |
| 662 | 34.42184 | -3.75893 | 22.09 | 0.77 | 1.943 | F |
| 574 | 34.42053 | -3.76141 | 22.46 | 1.43 | 1.931 | F |
| 514 | 34.42648 | -3.76251 | 22.69 | 1.32 | 1.935 | F |
| 483 | 34.41598 | -3.76315 | 22.12 | 1.38 | 1.930 | F |
| 631 | 34.41783 | -3.76003 | 22.79 | 0.45 | 1.933 | F |
| 607 | 34.44899 | -3.76058 | 23.36 | 0.89 | 1.941 | F |
| 598 | 34.41724 | -3.76130 | 23.59 | 1.42 | 1.939 | F |
| 623 | 34.45558 | -3.76030 | 22.96 | N/A | 1.948 | F |
| 5 | 34.43687 | -3.78074 | 22.74 | N/A | 1.931 | F |
| 954 | 34.45186 | -3.75109 | 23.57 | 0.32 | 1.927 | F |
| 181 | 34.42259 | -3.77175 | 23.65 | 0.33 | 1.936 | F |
| 750 | 34.45794 | -3.75725 | 22.29 | N/A | 1.923 | F |

**Extended Data Table 1 Measured properties of confirmed members of XLSSC 122 and z=1.93 structure**

Magnitudes are measured using the F140W filter and employ a Kron-type aperture. Colours are expressed as F105W-F140W magnitudes and are measured in 0.8-arcsecond circular apertures. The notes refer to gold (G) and silver (S) cluster members in addition to galaxies located in the foreground (F) structure; E refers to an emission line galaxy. Decl., declination. ID numbers are output from SExtractor.

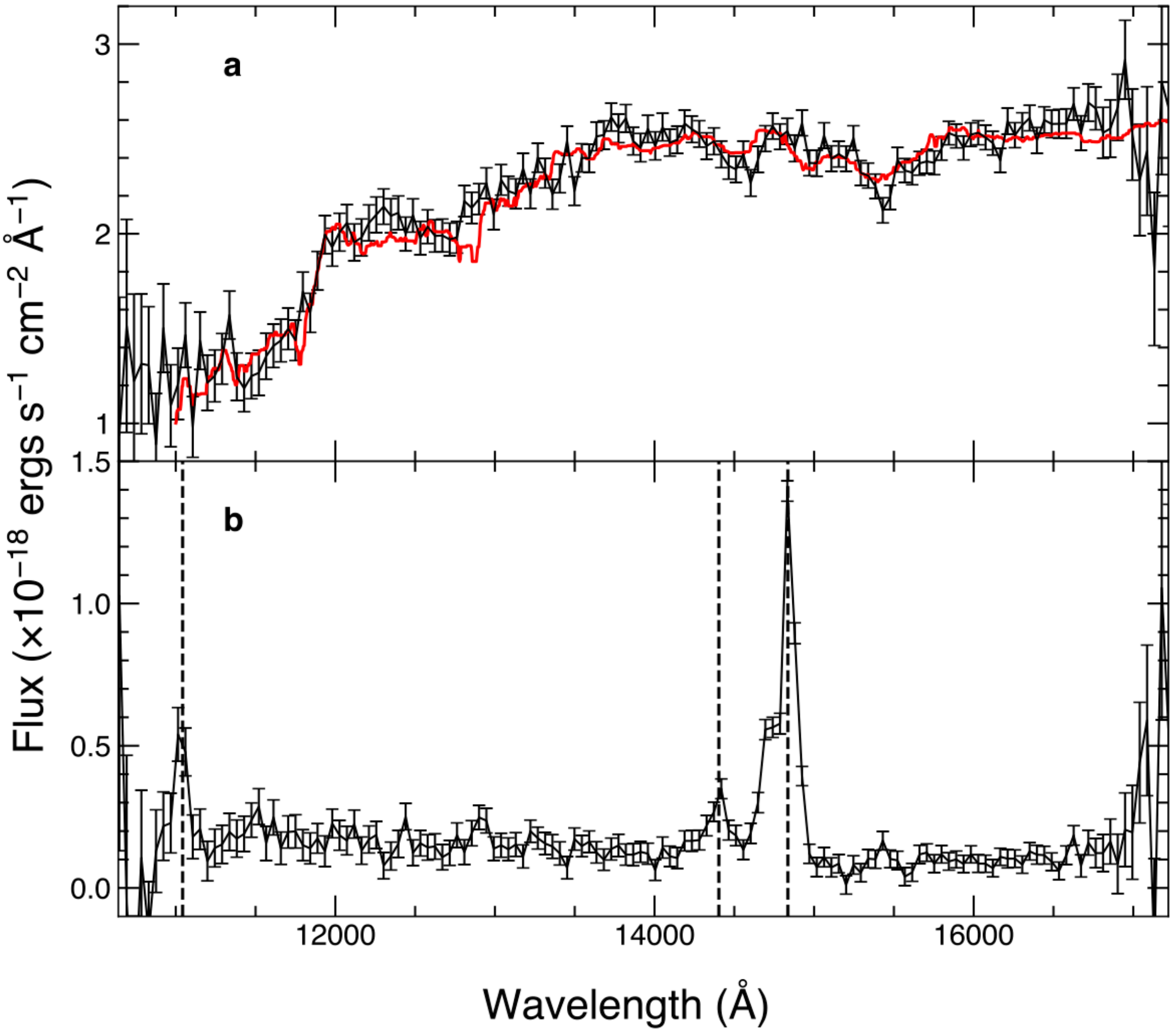

**Extended Data Fig. 1 Example spectra of two member galaxies of XLSSC 122. a**, The brightest cluster galaxy (ID 526) as the black line with error bars with the best-fitting, redshifted galaxy template shown in red (see Methods). **b**, A fainter cluster member with strong emission lines (ID 1141) as the black line with error bars. Error bars indicate the 1-sigma measurement uncertainty. The vertical dashed lines show the observed frame location of [O II] 3,727 Å, Hβ 4,861 Å and [O III] 5,007 Å at a redshift of 1.963.

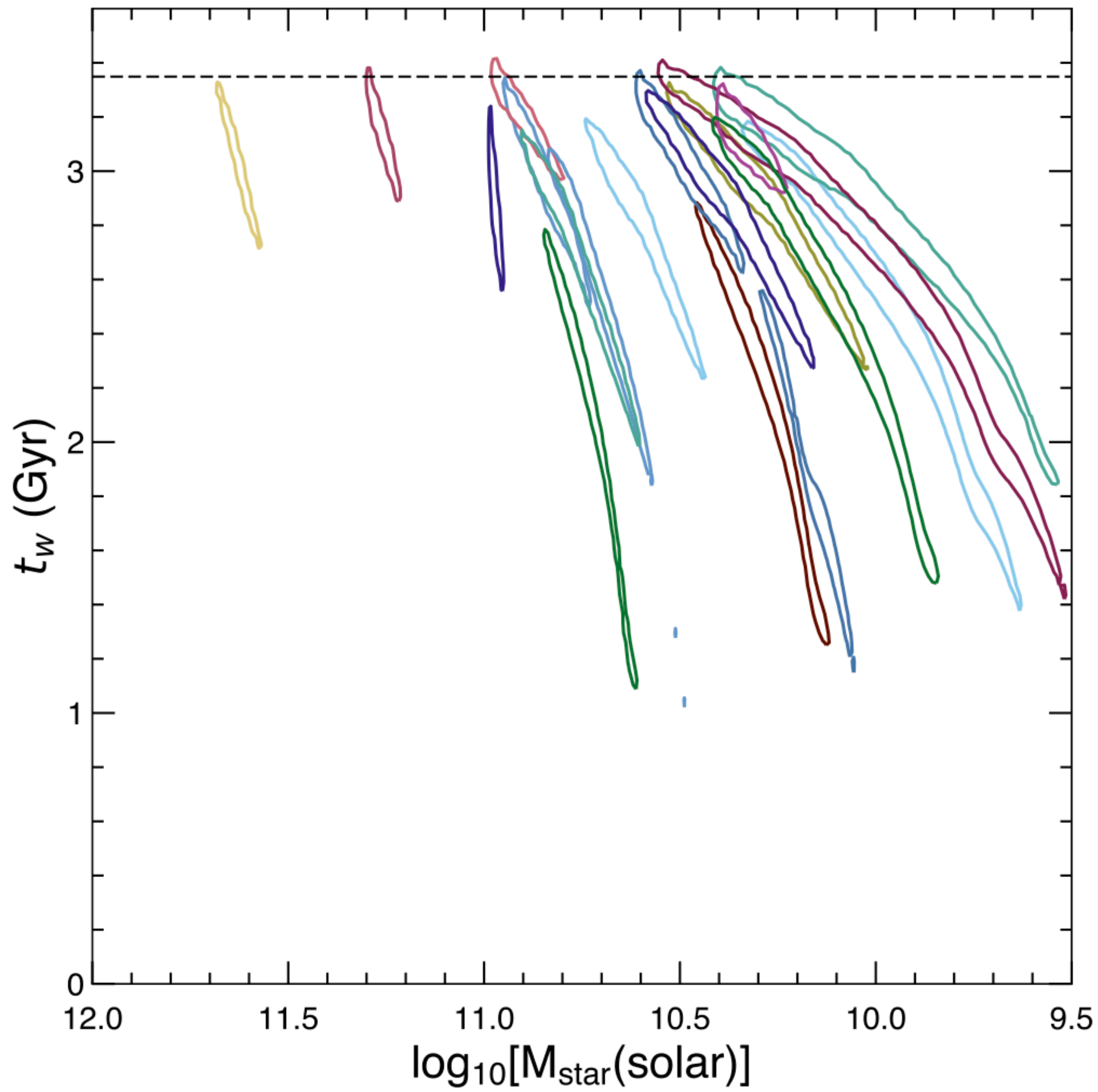

**Extended Data Fig. 2 The luminosity-weighted stellar age versus the mass of red sequence cluster member galaxies.** Posterior distributions in mean stellar age ($t_w$) and log stellar mass for the 19 'gold' members of the cluster red sequence. Only SED models assuming $A_V = 0.0$ are shown. Contours enclose 67% of the posterior probability for each galaxy. The horizontal dashed line indicates an age of 3.35 Gyr, that is, the age of the Universe at a redshift $z = 1.98$ in the assumed cosmological model.

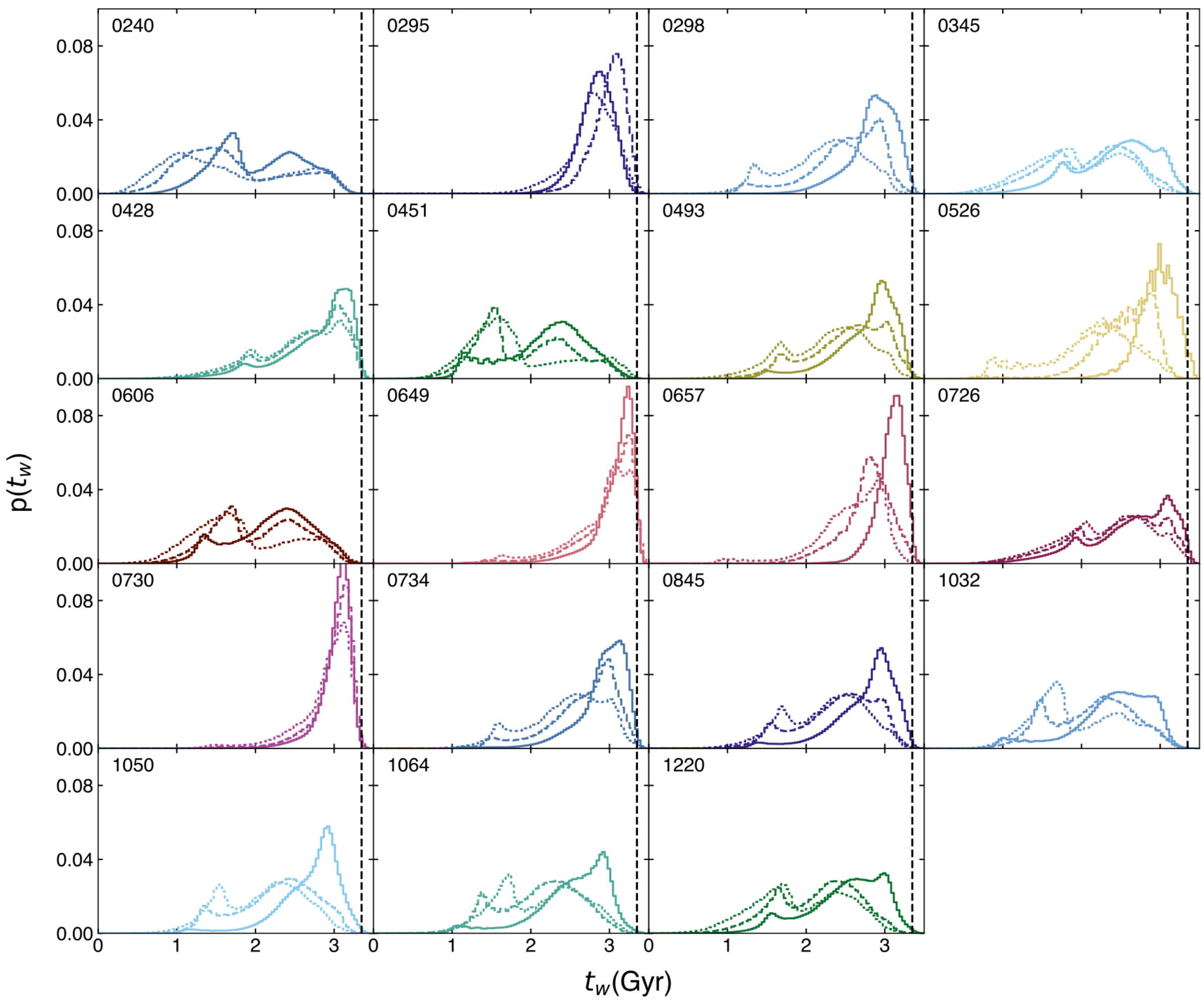

**Extended Data Fig. 3 The luminosity-weighted stellar age distributions for red-sequence cluster member galaxies.** Panels show posterior distributions in $t_{\mathrm{w}}$ for each 'gold' member galaxy of XLSSC 122, having marginalized over $M_{\mathrm{star}}$. For convenience, the same colour scheme is employed as in Extended Data Fig. 2. In each panel the solid, dashed and dotted curves display, respectively, SED models characterized by $A_{\mathrm{V}} = 0.0$, 0.3 and 0.5. The vertical dashed line in each panel indicates the age of the Universe at $z = 1.98$.